\documentclass[]{aastex631}

\usepackage{url}

\begin{document}

\title{Spatial coincidence between ultra-high energy cosmic rays and TeV gamma rays in the direction of GRB 980425/SN 1998bw}

\author{Nestor Mirabal}
\affiliation{Mail Code 661, Astroparticle Physics Laboratory, NASA Goddard Space Flight Center,
Greenbelt, MD 20771, USA}
\affiliation{University of Maryland, Baltimore County, MD 21250, USA}
\affiliation{Center for Research and Exploration in Space Science and Technology, NASA Goddard
Space Flight Center, Greenbelt, MD 20771}
\email{nestor.r.mirabalbarrios@nasa.gov}



\begin{abstract}
Gamma-ray bursts (GRBs) have long been suspected as possible ultra-high energy cosmic ray (UHECR) accelerators.
In this brief note, I report that GRB 980425/SN 1998bw falls within the region of interest (ROI) with the highest significance in an all-sky blind search for magnetically-induced effects in the arrival directions of UHECRs conducted by the Pierre Auger Collaboration with events detected up to to 2018 August 31. There is also report in the literature of delayed TeV emission in archival {\it Fermi}~LAT observations from the direction of GRB 980425/SN 1998bw. The combined probability that two distinct cosmic ray acceleration signatures in two different multimessenger experiments may appear at the same spatial location by chance is estimated to be between $1.62 \times 10^{-3}$ and 0.0157.  
\end{abstract}



\section{Introduction} 
GRBs are thought to be some of the most formidable candidates for UHECR acceleration \citep{waxman2,vietri}.
Motivated by recent discoveries consistent with secondary signatures of UHECR acceleration in GRB 980425/SN 1998bw \citep{2023JCAP...02..047M,2023JCAP...02..060M} and GRB 221009A 
\citep{2022arXiv221012855A,2023A&A...670L..12D,2023MNRAS.519L..85M}, 
a broader search of the literature was started to look for additional experimental evidence linking nearby GRBs to UHECR acceleration. In this note, I report that the nearby GRB 980425/SN 1998bw falls within the ROI which yields the highest significance in an all-sky blind search for magnetically induced signatures carried out by the
 Pierre Auger Collaboration  with events detected up to 2018 August 31 \citep{2020JCAP...06..017A}.
At a distance of 36.9 Mpc, GRB 980425/SN 1998bw lies inside the Greisen–Zatsepin–Kuzmin (GZK) horizon beyond which the propagation of $> 10^{20}$ eV protons is suppressed by proton interactions with the cosmic microwave background \citep{1966PhRvL..16..748G,1966JETPL...4...78Z}. 

\section{GRB 980425/SN 1998bw in the Pierre Auger data set}
\begin{figure}[t]
\plotone{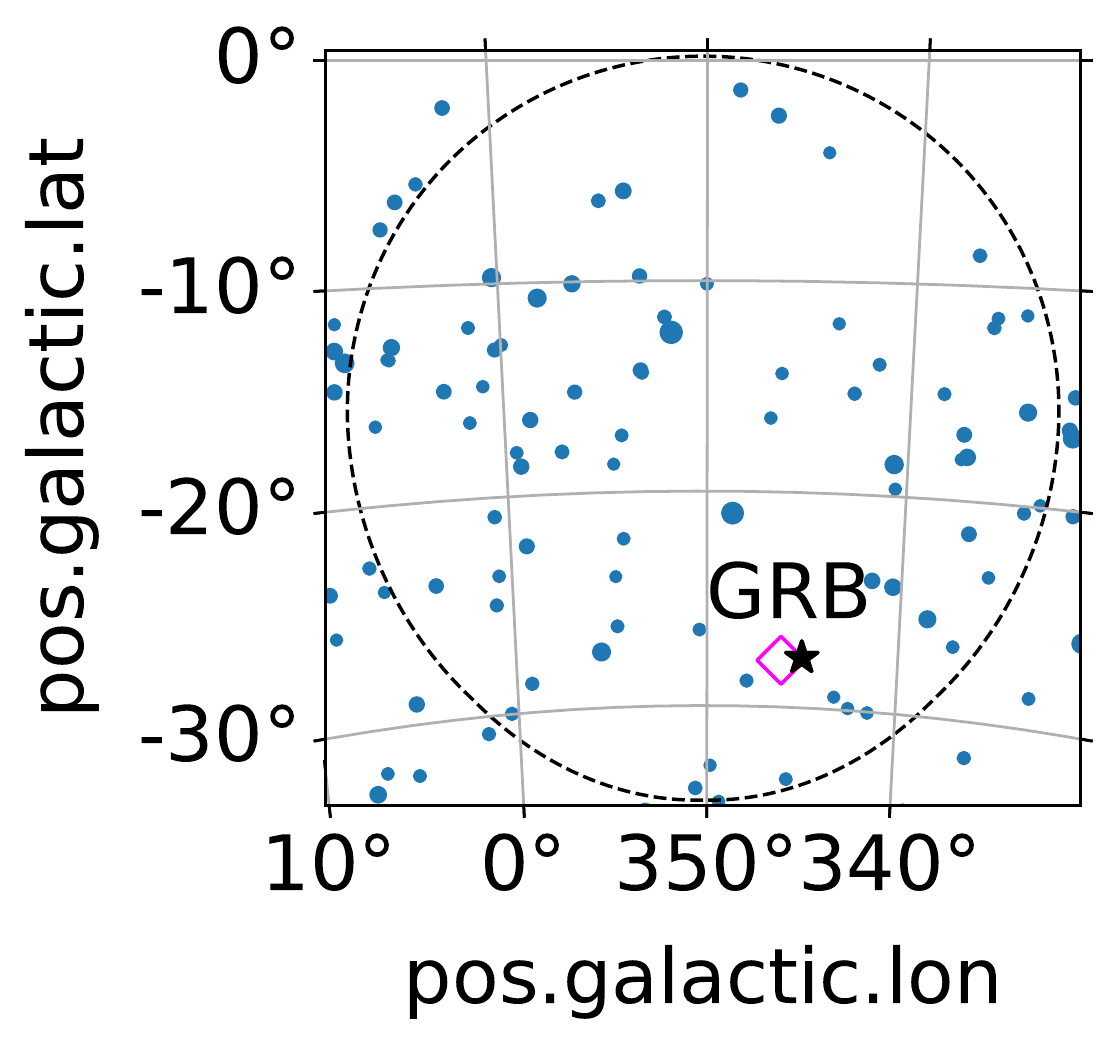}
\caption{Arrival directions of UHECRs (blue dots)
 above 32 EeV from January 1, 2004 to December 31, 2020 measured at the Pierre Auger Observatory \citep{2022ApJ...935..170A}. The image is centered on the highest significance ROI (dashed line) reported by \citet{2020JCAP...06..017A} The marker size is proportional to the individual UHECR energy from 32 EeV to 132.6 EeV. The star marks the position of GRB 980425/SN 1998bw. The magenta diamond marks the delayed TeV emission discussed in  \citet{2023JCAP...02..047M}.  
\label{fig1}}
\end{figure}

A search for signals of magnetically-induced effects in the arrival directions
of UHECRs was performed by the Pierre Auger Collaboration using events collected 
between 2004 January 1 and 2018 August 31, which
amounts to a total exposure of 101,900 km$^{2}$ sr yr \citep{2020JCAP...06..017A}. Two different methods were applied to the data. The first is the ``multiplet" method that looks for a correlation between the arrival directions of events and the inverse of their energy. The second method, which shall be referred to as the ``thrust" method, is a principal axis analysis designed to detect elongations in the arrival directions from a particular ROI  \citep{2020JCAP...06..017A}. During an all-sky blind search using the thrust method applied by the Pierre Auger Collaboration the most significant ROI ($p_{Auger}=0.25$) was found to have a thrust-ratio $T_{2}/T_{3} = 1.362$ for an energy threshold of 20 EeV centered at  ($\ell,b=350^{\circ}.2,-17^{\circ}$.0) 
with an ROI radius $r_{ROI}$ = 0.3 radians ($\sim 17.2 ^{\circ}$). The most significant Pierre Auger ROI includes and it is positionally consistent with the
GRB 980425/SN 1998bw coordinates ($\ell,b=344^{\circ}.99,-27^{\circ}$.72). Figure \ref{fig1} shows the GRB 980425/SN 1998bw position within the Pierre Auger ROI.

\clearpage
\section{Delayed Gamma-ray Emission from GRB 980425/SN 1998bw}
A dedicated search for delayed TeV emission from nearby GRBs found possible evidence
of $>1$ TeV emission approximately 0.96 degrees away from the position of GRB 980425/SN 1998bw \citep{2023JCAP...02..047M}. 
The delayed TeV emission has been interpreted as secondary emission from UHECRs initially accelerated by GRB 980425/SN 1998bw \citep{2022MNRAS.511.5823R} and deflected by galactic/intergalactic magnetic fields (IGMF). As UHECRs propagate, 
nucleons interact with the microwave background to generate delayed secondary GeV--TeV photons. 
 The Poisson probability of observing one or more TeV photons within 0.96 degrees of GRB 980425/SN 1998bw
has been reported as 
$p_{Fermi} = 0.0088$ for {\it Fermi}-LAT data collected
between  2008 August 4 and 2022 September 20 \citep{2023JCAP...02..047M}. 

\section{Combined Probability}
The initial calculation of the combined probability $p$ 
assuming two independent observations with probabilities $p_{Auger}$ and
$p_{Fermi}$ respectively is given by the expression obtained by \citet{2006smep.book.....J} 

\begin{equation}
p = p_{Auger} p_{Fermi}(1 - ln(p_{Auger} p_{Fermi})).
\end{equation}

Using this prescription, one finds that $p = 0.0157$. 

A second type of calculation provides a more direct overlap comparison through a corrected probability $p_{Auger'}$ for picking a single Auger ROI over the rest of the observable Auger sky. In this case, 
the ROI with the highest significance detected by the Pierre Auger Observatory corresponds to a fraction of the 3.4$\pi$ steradians covered with the Pierre Auger Observatory in any season \citep{2022ApJ...935..170A}. The probability of picking this single circular ROI with a radius $r = 0.3$ radians over the entire coverage region is $p_{Auger'} = \frac{0.3^{2}}{3.4} = 0.026$. As before, one can recalculate the alternative combined probability   $p'$ using a similar prescription 

\begin{equation}
p' = p_{Auger'} p_{Fermi}(1 - ln(p_{Auger'} p_{Fermi})),
\end{equation}

\noindent
yielding $p' = 1.62 \times 10^{-3}$. Since there are two different values for the combined probability, the actual combined probability is within the probability range from $1.62 \times 10^{-3}$ to 0.0157. 

\section{Implications and Future Work}
While the combined probability range does not unequivocally link GRBs and UHECRs, the fact that two distinct signatures in two independent multimessenger facilities spatially overlap in the direction of the nearest GRB ever detected seems to strengthen the suggestion that GRB 980425/SN 1998bw was a transient source of UHECRs within the GZK horizon. 
This would be consistent with the theoretical proposal that UHECRs are accelerated in nearby GRBs \citep{waxman2,vietri}. 
As pointed out in \citet{2023JCAP...02..047M,2023JCAP...02..060M}, it is possible that a more significant UHECR excess will emerge incrementally from the direction of GRB 980425/SN 1998bw. It is worth noting that   
the multiplet/thrust ratio probabilities discussed in this note only included Pierre Auger events up to 2018 August 31. With almost 5 more years of Pierre Auger Observatory events already collected, an improved combined significance might be obtained. Additional targeted searches for magnetically-induced signatures using the multiplet and thrust methods should be performed in and around the GRB 980425/SN 1998bw localization. Efforts should also be made to effectively model magnetically-induced signatures assuming various heavy nuclei and Galactic magnetic field deflection configurations.


\begin{acknowledgments}
The material is based upon work supported by NASA under award number 80GSFC21M0002.
All data used in this note are publicly available. I acknowledge informative correspondence with Geraldina Golup and Ralph Engel.
\end{acknowledgments}

\software{Astropy \citep{astropy:2022}, \url{https://git.ligo.org/leo-singer/ligo.skymap} }

\bibliography{sample631}{}
\bibliographystyle{aasjournal}



\end{document}